\documentclass[a4paper,11pt]{article} 

\usepackage[latin1]{inputenc} 
\usepackage{amssymb}
\usepackage{amsmath}
\usepackage{tikz}

\begin{document}
\setlength{\unitlength}{1mm}
\renewcommand{\baselinestretch}{1.1}
\renewcommand{\i}{i}
\allowdisplaybreaks[1]

\title{A Theory of Inertia Based on Mach's Principle}
\author{Volkmar PUTZ}
\date{}
\maketitle
\begin{center}  
University College of Teacher Education Vienna (PH Wien)\\
Grenzackerstrasse 18, A-1100 Vienna, Austria\\
e-mail: volkmar.putz@phwien.ac.at
\end{center}

\begin{abstract}
A non-relativistic theory of inertia based on Mach's principle is presented as has been envisaged but not achieved by Ernst Mach in 1872. Central feature is a space-dependent, anisotropic, symmetric inert mass tensor.
\end{abstract}

\section{Introduction}

In 1872 Ernst Mach formulated his famous principle about the inert mass of a particle, which he assumed to be induced somehow by the presence of other masses in the universe \cite{Mach}. Thereby, inertia is the result of gravitational forces. Unfortunately, he never formulated this as a physical theory. 

In this paper, I want to present a theory consistent with Mach's principle. The theory is non-relativistic. Hence it is more of a historical value, giving what Mach did not (or could not) deliver nearly 150 years ago, when Mach's ideas were not yet overgrown by Einstein's theories (which, contrary to Einstein's hopes and struggles, do not fully incorporate the Mach principle).

A central aspect of the theory is the anisotropy of the inert mass, which therefore has to become a tensor. In the 1950s, experiments in search of an anisotropy of inertia have been performed with negative result \cite{Hughes}. Later, these experiments have been recognized as not suitable, ignoring the fact that locally not only the test particle but all masses (including those of the experimental setup) exhibit the same anisotropy of inertia \cite{Anderson}. Different experiments where this effect does not matter are still to be performed.

\section{Assumptions}

We start with the following assumptions, which are all very simple and plausible:

\vspace{0.1cm}\noindent
{\bf Assumption 1:} A single point mass in an otherwise empty universe has no defined movement, {\it i.e.}~no inertia, no acceleration, no momentum or kinetic energy.

\vspace{0.1cm}\noindent
{\bf Assumption 2:} Two point masses in an otherwise empty universe can have a defined movement only along their connecting line. Any movement at constant distance between the masses cannot be defined. This comprises rotation about each other (no tangential inertia) and collective translation. No kinetic energy or momentum arises due to such movement: Inertia, acceleration, momentum or kinetic energy can only be defined with respect to increasing/reducing the distance between the masses, {\it i.e.}~for radial movement. We would like to have a theory where the sign of this radial movement does not matter.

\vspace{0.1cm}\noindent
{\bf Assumption 3:} Since the inertia of a test particle is 'created' by the gravitational forces of other masses, it will depend on the direction. Therefore the angle between the direction of the test particle movement and the connecting line to any inducing mass will be crucial.

\vspace{0.1cm}\noindent
{\bf Assumption 4:} The size of the induced inertia should decrease with increasing distance $r$ between test particle and inducing mass as $1/r^a$ with $0<a<2$. Here $0 < a$ guarantees vanishing inertia at infinite distance (which, of course, is an assumption itself). $a \geq 2$ would lead to large effects perceivable in daily life, since anisotropy of inertia would balance or even excel the effects of Newton's gravity. In fact, $a = 1$ is a standard assumption in literature, see {\it e.g.}~\cite{TrederRel}, where, however, isotropy of inertia is postulated.

\vspace{0.1cm}\noindent
{\bf Assumption 5:} The inertia of a test particle within a hollow sphere of constant mass density is isotropic and constant. If the mass distribution is largly isotropic within our universe, this assumption guarantees that this theory is very similar to our daily experience.

\section{Theory}

A definition of the kinetic energy of a test particle with mass $m_0$ within a universe with other (inertia inducing) masses $m^\alpha$, which is consistent with all assumptions above is given by:
\begin{eqnarray}\label{energy}
E_{kin} = \frac 12\sum_\alpha v_i^\alpha M_{ij}^\alpha v_j^\alpha\ ,\qquad\\
M_{ij}^\alpha = M_{ji}^\alpha = \gamma m_0 m^\alpha\frac{r_i^\alpha r_j^\alpha}{\vert \vec r^{\,\alpha}\vert^3}\ . 
\end{eqnarray}
$\gamma$ is a natural constant, presumably connected with Newton's gravitational constant. $\vec r^{\,\alpha} = \vec x_0 - \vec x{\,^\alpha}$ is the distance between $m_0$ and $m^\alpha$, with $\vec x_0$ and $\vec x^{\,\alpha}$ the absolute positions of $m_0$ and $m^\alpha$ in space, respectively. Similarly, $\vec v^{\,\alpha} = \dot{\vec x}_0 - \dot{\vec x}^{\,\alpha}$ is the {\it relative} (not only radial!) velocity between $m_0$ and $m^\alpha$ (as will be the relative acceleration $\dot{\vec v}^{\,\alpha} = \vec a^{\,\alpha}$ later on). The sum is over all masses $m^\alpha$ in the universe with the exception of $m_0$.

The corresponding components of the momentum and force of the test particle $m_0$ are then given by
\begin{eqnarray}
p_i &=& \frac{\partial E_{kin}}{\partial  (\vec v_0)_i} = \sum_\alpha M_{ij}^\alpha v_j^\alpha\ ,\\
F_i &=& \frac{dp_i}{dt} = \sum_\alpha \left(\dot M_{ij}^\alpha v_j^\alpha + M_{ij}^\alpha a_j^\alpha\right)\ .
\end{eqnarray}
If and only if we are in a coordinate system where {\it all} inducing masses $m^\alpha$ are at rest, we have $\vec v^{\,\alpha} = \vec v_0$ and the formulae simplify to
\begin{eqnarray}
E_{kin} = \frac12\vec {v_0}^{\,T}\, \overleftrightarrow{M}\,\vec v_0\ ,\quad 
\vec p = \overleftrightarrow{M}\,\vec v_0\ , \quad
\overleftrightarrow{M} = \sum_\alpha m^\alpha \frac{{\vec r}^{\,\alpha}\ ({\vec r}^{\,\alpha})^T
}{\vert {\vec r^{\,\alpha}}\vert^3}
\end{eqnarray}

\section{Check of Assumptions}

Let us now investigate how this definitions fullfill assumptions 1-5. For this, we do not consider Newtonian  gravitation beyond induction of inertia.

\vspace{0.1cm}\noindent
{\bf Assumption 1:} Since there exists only $m_0$ and not one single $m^\alpha$, the $M_{ij}^\alpha$ are all zero. There is no kinetic energy and no momentum and also no force.

\vspace{0.1cm}\noindent
{\bf Assumption 2:} For translations at constant distance $\vec v^{\,\alpha} = 0$. For (possibly non-uniform) rotations of a mass $m_0$ with constant radius $R$ around a single point mass $m^\alpha$ (or vice versa), we assume $\vec x^{\,\alpha} = (0,0,0) \Rightarrow \vec r^{\,\alpha} = \vec x_0 - \vec 0 = (R\cos \omega t, R\sin \omega t, 0)$, with $\omega$ possibly time dependent. Thus we have
\begin{eqnarray}
\overleftrightarrow{M} &=& \frac{\gamma m_0 m^\alpha}R\left(\begin{array}{ccc} \cos^2\omega t & \sin\omega t\cos\omega t & 0\\ \sin\omega t\cos\omega t &\sin^2\omega t & 0\\ 0 & 0 & 0\end{array}\right)\ ,\\
\vec v^\alpha &=& R(\dot\omega t +\omega)\left(\begin{array}{c}-\sin \omega t\\ \cos \omega t\\ 0\end{array}\right)\ .
\end{eqnarray}
With our definitions above one sees easily that $\vec p = \vec 0,\ E = 0$ and again there is no force (this is also true if $m^\alpha$ and $m_0$ rotate synchronously around some other rotational center, {\it e.g.}~both being fixed onto one single watch hand). Especially, there are no centripetal or centrifugal forces. (Therefore, the gravitational attraction between the two masses will eventually lead to collision, regardless if there is a rotation or not.)

\vspace{0.1cm}\noindent
{\bf Assumption 3:} Anisotropy of inertia is realized in our definition. As an illustration let us consider a simple experiment involving three masses. $m_0$ is at rest in $(0,0,0)$. We have two masses $m^\alpha$, with mass $m^1 = M$ at $(0,-R,0)$ and mass $m^2 = 2M$ at $(-R,0,0$).

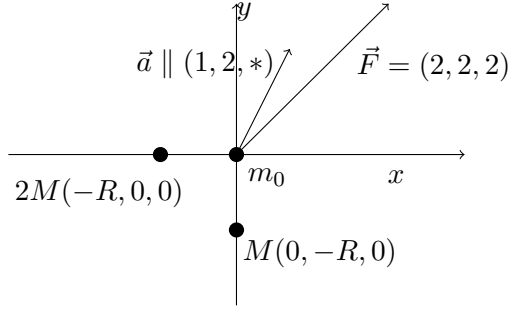
\begin{center}
\begin{tikzpicture}
\draw(1.2,-0.5) node{$2M(-R,0,0)$};
\draw(4.1,-1.3) node{$M(0,-R,0)$};
\draw(3.4,-0.3) node{$m_0$};
\draw(5.6,1.2) node{$\vec F = (2,2,2)$};
\draw(2.6,1.2) node{$\vec a \parallel (1,2,*)$};
\draw(3.1,1.9) node{$y$};
\draw(5.1,-0.3) node{$x$};
\draw[->] (0,0) -- +(6, 0);
\draw[->] (3,-2) -- +(0, 4);
\fill(2,0) circle (0.1);
\fill(3,0) circle (0.1);
\draw[->](3,0) -- +(2, 2);
\draw[->](3,0) -- +(0.7, 1.4);
\fill(3,-1) circle (0.1);
\end{tikzpicture}
\end{center}

\noindent
After a short calculation, we find (with $\vec p_0 = 0$)
\begin{eqnarray}
\overleftrightarrow{M} &=& \frac {\gamma m_0 M}R \left(\begin{array}{ccc} 2 & 0 & 0\\ 0 & 1 & 0 \\ 0 & 0 & 0\end{array}\right)\ , \qquad \vec F = \overleftrightarrow{M} \vec a\ .
\end{eqnarray}
If a force $\vec F = (2,2,2)$ is applied, this will lead to an acceleration of $\vec a = (1,2,*)$, which is clearly not the same direction as $\vec F$. The * signals the not defined acceleration in the here inertia-free $z$-direction.

If $\vec p_0 \neq \vec 0$ at the beginning, things become more complicated, since the change of $\vec r^{\,1}$ and $\vec r^{\,2}$ (the position of $m_0$ relative to the two other masses) gives extra terms leading to $\vec a \nparallel (1,2,*)$.

\vspace{0.1cm}\noindent 
{\bf Assumption 4:} One sees easily that the contribution of each mass $m^\alpha$ to $\overleftrightarrow{M}$ is proportional (besides the dependence on direction) to $1/\vert \vec r^{\,\alpha}\vert$.

\vspace{0.1cm}\noindent
All these considerations can easily be generalized to systems with more masses $m^\alpha$, where one also has {\it e.g.}~the absence of kinetic energy and momentum if all masses move uniformly or the absence of centrifugal forces if all masses rotate uniformely without change of relative distances. This corresponds perfectly to Mach's famous bucket-gedankenexperiment.

\vspace{0.1cm}\noindent
{\bf Assumption 5} is perhaps the most crucial and definitely the most complicated to prove. In the next section, we will give the detailed calculation (\ref{M}). The result for the inert mass tensor of a test particle $m_0$ (resting or moving with constant velocity relative to the sphere) within a thin sphere of radius $R$ and constant area mass density $\sigma$ (and hence $4\pi R^2 \sigma$ its overall mass) is:

\begin{eqnarray}
\overleftrightarrow{M} &=& \frac\gamma3\frac{4\pi R^2\sigma}Rm_0 \left(\begin{array}{ccc} 1 & 0 & 0\\ 0 & 1 & 0 \\ 0 & 0 & 1\end{array}\right)\ .
\end{eqnarray}
This corresponds to a constant, isotropic inertia independent of the spatial position (as long as within the sphere) and would therefore be the physical setting for the emergence of usual Newtonian inertia.

\section{Calculation of M within a sphere}

We start with a test particle at the origin of our coordinate system. This particle is within a thin sphere of constant area mass density $\sigma$ with center $(0|0|-a)$ and radius $R$ ($R > a$). 

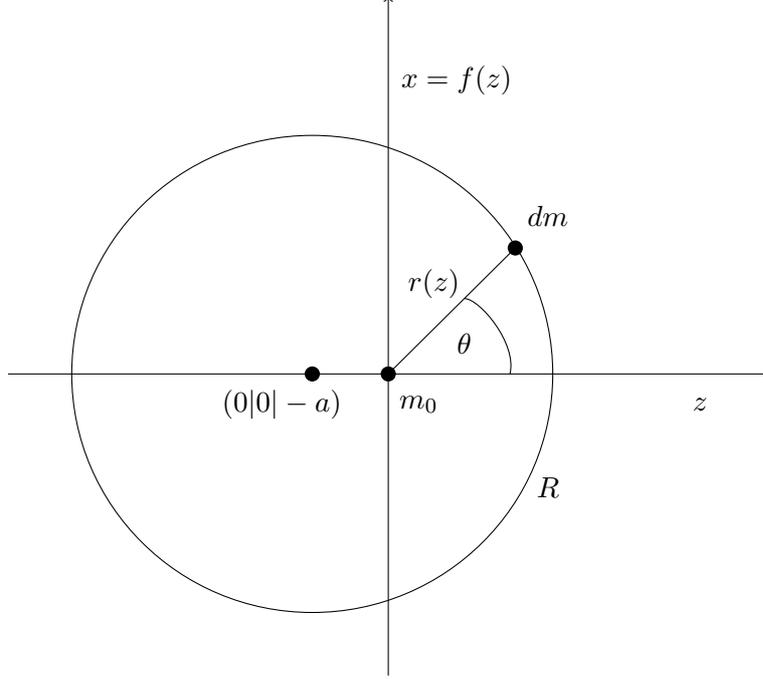
\begin{tikzpicture}
\draw(3.6,-0.4) node{$(0|0|-a)$};
\draw(5.4,-0.4) node{$m_0$};
\draw(6,0.4) node{$\theta$};
\draw(5.6,1.2) node{$r(z)$};
\draw(7.1,2.1) node{$dm$};
\draw(7.1,-1.5) node{$R$};
\draw(5.9,3.9) node{$x = f(z)$};
\draw(9.1,-0.4) node{$z$};
\draw[->] (0,0) -- +(10, 0);
\draw[->] (5,-4) -- +(0, 9);
\draw (4,0) circle (90pt);
\fill(4,0) circle (0.1);
\fill(5,0) circle (0.1);
\draw (5,0) -- +(1.67, 1.67);
\fill(6.67,1.67) circle (0.1);
\draw (6,1) .. controls (6.2,1) and (6.7,0.4) .. (6.6, 0);
\end{tikzpicture}

\noindent
The mass element $dm$ of the sphere is at point $(z, x = f(z), 0)$, where
\begin{eqnarray*}
f(z) &=& \sqrt{R^2- (z+a)^2}\ ,\\
f'(z) &=& - \frac{z+a}{\sqrt{R^2- (z+a)^2}}\ ,\\
f(z)\sqrt{1+f'^2(z)} &=& \sqrt{R^2- (z+a)^2}\sqrt{1+ \frac{(z+a)^2}{R^2- (z+a)^2}}
= R\ ,\\
r(z) &=& \sqrt{z^2 + f^2(z)} = \sqrt{R^2 - a^2 - 2az} \ .
\end{eqnarray*}
For later use, we want to calculate the gravitational potential $\Phi_{in}$ within the sphere:
\begin{eqnarray}\label{Treder}
m_0\Phi_{in} &=& \int_0^{2\pi} d\varphi\int_{-R-a}^{R-a} dz \ \frac{m_0\gamma\sigma}{r(z)} \underbrace{f(z)\sqrt{1+f'^2(z)}}_{R}\\\nonumber
&=& 2\pi R m_0\gamma\sigma \int_{-R-a}^{R-a} \frac{dz}{\sqrt{R^2-a^2-2az}} \\\nonumber
&=& 
2\pi R m_0\gamma\sigma\ \frac{-1}a \sqrt{R^2-a^2-2az}\Big\vert_{-R-a}^{R-a} \\\nonumber
&=& \frac{2\pi R m_0\gamma\sigma}a \Big(\underbrace{\sqrt{R^2-a^2+2aR+2a^2}}_{R+a} - \underbrace{\sqrt{R^2-a^2-2aR+2a^2}}_{R-a}\Big)\\\nonumber
&=& 4\pi R m_0\gamma\sigma\ .  
\end{eqnarray}
Of course, this result is not at all surprising: The gravitational potential within a hollow sphere of homogeneous mass density is constant, {\it i.e.}~independent of the position.

We now want to calculate $\overleftrightarrow{M}$ at the coordinate origin (where $m_0$ is located):
\begin{eqnarray*}
&&\overleftrightarrow{M} = \int_{\mathrm{sphere}} d^3r\ m_0\gamma\sigma\ \frac{\vec r\ \vec r^{\,T}}{\vert \vec r\vert ^3}
\\&&=   \int_0^{2\pi} d\varphi\int_{-R-a}^{R-a} dz\ \frac{m_0\gamma\sigma}{r^3(z)}  \underbrace{f(z)\sqrt{1+f'^2(z)}}_{R}
\left(\begin{array}{ccc} x^2 & xy & xz \\ yx & y^2 & yz \\ zx & zy & z^2\end{array}\right)\ .
\end{eqnarray*}
With $(x,y,z) = (r(z)\sin\theta\cos\phi,\ r(z)\sin\theta\sin\phi,\ r(z)\cos\theta)$ we see immediately that after the $\phi$-integration the off-diagonal terms are all zero and the diagonal terms yield $2\pi$ for the $z^2$-term and $2\pi/2$ for the other two:
\begin{eqnarray*}
\overleftrightarrow{M} &=& 
2\pi R m_0\gamma\sigma\int_{-R-a}^{R-a} dz  \frac{1}{r(z)}
\left(\begin{array}{ccc} \frac12\sin^2\theta  & 0 & 0 \\ 0 & \frac12\sin^2\theta & 0 \\ 0 & 0 &\cos^2\theta\end{array}\right)\ .
\end{eqnarray*}
We start with the $zz$-term  proportional to $\cos^2\theta$. We rewrite:
\begin{eqnarray*}
\cos^2 \theta &=& \left(\frac{z}{r(z)}\right)^2 = \frac{z^2}{R^2-a^2-2az}
\end{eqnarray*}
and get:
\begin{eqnarray*}
M_{zz} &=& 2\pi Rm_0\gamma\sigma\ I_z = 2\pi Rm_0\gamma\sigma \int_{-R-a}^{R-a} dz  \frac{z^2}{(R^2-a^2-2az)^{3/2}}\ . 
\end{eqnarray*}
Here we use the well known formula \cite{Bartsch}
\begin{eqnarray*}
&&\int dz \frac{z^2}{(Az+B)^{3/2}} \\ &&= \frac2{A^3}\Big(\frac13(Az+B)^{3/2} - 2B(Az+B)^{1/2}
- B^2(Az+B)^{-1/2}\Big)\ ,
\end{eqnarray*}
where in our case $A = -2a$ and $B = R^2-a^2$ (note that the denominator is $r^3(z) > 0$):
\begin{eqnarray*}
\int_{-R-a}^{R-a}\frac{z^2}{(R^2-a^2 - 2az)^{3/2}}&=& -\frac1{4a^3}\Big(\frac13(R^2-a^2 - 2az)^{3/2}
\\ &&- 2(R^2-a^2)(R^2-a^2 - 2az)^{1/2} \\&& - (R^2-a^2)^2(R^2-a^2 - 2az)^{-1/2}\Big)\Big\vert_{-R-a}^{R-a}\ .
\end{eqnarray*}
We notice again
\begin{eqnarray*}
&&(R^2-a^2-2a(R-a))^{1/2} = R-a\ ,\\
&&(R^2-a^2+2a(R+a))^{1/2} = R+a\ ,
\end{eqnarray*}
and find:
\begin{eqnarray*}
I_z &=& -\frac1{4a^3}\Big(\frac13(R-a)^3 - 2(R^2-a^2)(R-a) - \frac{(R^2-a^2)^2}{R-a}\Big)\\
&&+ \frac1{4a^3}\Big(\frac13(R+a)^3 - 2(R^2-a^2)(R+a) - \frac{(R^2-a^2)^2}{R+a}\Big)\\
&=& \frac1{4a^3}\cdot \frac83 a^3 = \frac23\ ,\\
M_{zz} &=& 2\pi Rm_0\gamma\sigma\  I_z = \frac13\cdot 4\pi Rm_0\gamma\sigma\ .
\end{eqnarray*}
This is exactly one third of what we had before (\ref{Treder}), independent of the test particle's position within the sphere.

For the $xx$- and $yy$-component, the integration is easy using $\frac12{\sin^2\theta} = \frac12(1-\cos^2\theta)$ with the results from above:
\begin{eqnarray*}
2\pi Rm_0\gamma\sigma I_{x,y} = \frac12(1-1/3)\cdot 4\pi Rm_0\gamma\sigma 
= \frac13\cdot 4\pi Rm_0\gamma\sigma = I_z\ .
\end{eqnarray*}
Again, it is exactly $1/3$ of what we had before. In sum, we end up with
\begin{eqnarray}\label{M}
\overleftrightarrow{M} &=& \frac\gamma3\frac{4\pi R^2\sigma}Rm_0 \left(\begin{array}{ccc} 1 & 0 & 0\\ 0 & 1 & 0 \\ 0 & 0 & 1\end{array}\right)\ .
\end{eqnarray}
{\it Remark:} In this calculation, one sees that the contribution of a mass element $dm$ to the inertia of a particle $m_0$ at rest (experiencing an acceleration) is proportional to $\cos^2\alpha$, where $\alpha$ is the angle between the line connecting $m_0$ and $dm$ and the direction of the acceleration. So if the angle is $90^\circ$, there is no contribution of $dm$ to the inertia, whereas the contribution is maximal if both directions are in parallel.

\section{Conclusion}

In this paper, a non-relativistic theory of inertia based on Mach's principle was presented. The theory is by no means intended to replace Einstein's general theory of relativity (GRT), {e.g.}~it will not explain perihelion precession of planets since the gravitation of the central star (at this scale nearly a point mass) has barely any influence on the tangential inertia of surrounding objects. However, this theory may come to life at the scale of galaxies and their movement. In any case, the next step will be to (try to) incorporate this theory into GRT. 

The most crucial task would be a proper testing of inert (an)isotropy using experiments which do not rely only on masses ({\it i.e.}~gravitation and inertia but also {\it e.g.}~electric fields. Since induced inertia and Newtonian gravitation always come together, interference of both effects may complicate things.

\end{document}